\documentstyle[12pt]{article}
\begin{document}

\noindent
{\hfill\it Astronomical Society of Australia Newsletter, 2002, 26, 4}

\bigskip
\bigskip
\centerline
{\Large \bf The 2001 Astronomy Job Market}

\centerline{Summary of a Talk Presented at the 2001 Harley Wood Winter School}

\centerline{June 2001, Lorne, Victoria, Australia}

\bigskip
\centerline{\large Brad K. Gibson}

\smallskip
\noindent
{\centerline{Centre for Astrophysics \& Supercomputing}

{\centerline{Swinburne University, Hawthorn, Victoria, Australia, 3122}

\bigskip
\bigskip

Over the 12 month period spanning July 2000 to June 2001, a survey of the
American Astronomical Society monthly Job Registers showed that $\sim$300
Postdoctoral Fellowships (PDFs) in astronomy were available.  To put that into
perspective, there are $\sim$250 new astronomy/astrophysics PhDs 
awarded each year
($\sim$1/2 of which originate from US degree-granting institutes).   
Historically,
there are $\sim$70 new permanent positions on the market per year.  
One bottom line
is that $\sim$3/4 of 
PhD graduates will land a PDF position (i.e. after discounting
those who voluntarily leave the field, virtually all graduates who want a PDF
will find one).  The overproduction rate of PhDs-to-permanent positions remains
at $\sim$3.

The number of permanent positions advertised in this 12 month period was
up somewhat: 83.5 ``permanent'' positions were advertised in the AAS Job
Register, 69.5 of which were based in the US, and 47.5 of the 83.5 faculty
positions were ``targeted'' (in the sense that a specific area of
astrophysics was required). The breakdown makes for interesting reading
and should be brought to the attention of prospective PhD students: 48\% of
the positions were targeting `theory', 32\% were targeting `observation',
and 20\% were targeting `instrumentation'.  In terms of research areas, 56\%
were looking for cosmologists of one sort or another, 25\% were looking for
planetary scientists, 11\% were looking for high energy astrophysicists,
and 8\% were looking for stellar astronomers.

Some peculiarities to the Australian community were highlighted,
including the following: (1) Australian institutes produce astronomy
PhDs at a rate in excess of that encountered in other countries (only
10\% of Australian graduates end up with permanent positions in
astronomy in Australia); (2) to their detriment, Australian students
were reluctant to change institutes between their BSc and PhD; (3) a
perception exists (in some corners) overseas that Australian students
undertake too much observing during their PhD (``excellent observers,
but not all that well-trained in astrophysics'').

In terms of a so-called ``recipe for success'' for prospective
astronomy PhD students, the following was suggested: (1)
investigate the past history of both the school and supervisor
who interests you (e.g. how have their past students fared?
What is their grant history like?  What are the opportunities
for external \& international collaborative links?); (2) know
what's hot (both in the near- and long-term -  e.g.
computational and/or theory, cosmology, instrumentation,
planetary, astrobiology); (3) get experience writing grants; (4)
be sure to network; (5) avoid excessive observing; (6) be wary
of undertaking PhDs within very large teams; (7) for Australian
students, move institutes for your PhD and PDF, and be prepared
to emigrate; (8) be aware of the competition (most folks have
$\sim$10 papers by the end of their first PDF position). 

\end{document}